\documentclass[review]{elsarticle}

\usepackage{hyperref}
\usepackage{rotating}

\journal{Journal of \LaTeX\ Templates}

\begin{document}

\begin{frontmatter}

\title{Review of the 1$^{st}$ EUV Light Sources Code Comparison Workshop}

\author[1,2]{J. Sheil\corref{cor1}}
\ead{j.sheil@arcnl.nl}
\address[1]{Advanced Research Center for Nanolithography, Science Park~106, 1098~XG Amsterdam, The Netherlands}
\cortext[cor1]{Corresponding Author} 
\address[2]{Department of Physics and Astronomy, and LaserLaB, Vrije Universiteit Amsterdam, De Boelelaan 1081, 1081 HV Amsterdam, The Netherlands}

\author[2]{O. O. Versolato$^{\mathrm{a},}$}

\author[3]{V. Bakshi}
\address[3]{EUV Litho, Inc., 10202 Wommack Road, Austin, Texas 78748, United States of America}

\author[4]{H. A. Scott}
\address[4]{Lawrence Livermore National Laboratory, Livermore, California 94551, United States of America}









\begin{abstract}
We review the results of the 1$^{st}$ Extreme Ultraviolet (EUV) Light Sources Code Comparison Workshop, which was held online on 3$^{rd}$ November 2020. The goal of this workshop was to provide a platform for specialists in EUV light source plasma modeling to benchmark and validate their numerical codes using well-defined case studies. Eight institutions spanning four countries contributed data to the workshop. Two topics were addressed, namely (i) the atomic kinetics and radiative properties of tin plasmas under EUV-generating conditions and (ii) laser absorption in a fully ionized, one-dimensional hydrogen plasma. In this paper, we summarize the key findings of the workshop and outline plans for future iterations of this code comparison activity.
\end{abstract}


\end{frontmatter}


\section{Introduction}

The generation, diagnosis and characterization of plasmas formed on mid- to high-$Z$ elements through intense laser irradiation nowadays constitutes a large component of high energy density physics (HEDP) research. A prominent example of this is indirect-drive inertial confinement fusion (ICF) \cite{Lindl1995,Kline2019,Zylstra2022}, where high-energy x-ray radiation generated in a multi-laser-driven gold ($ Z = 79 $) holhraum plasma drives the compression of a deuterium-tritium-filled capsule to fusion conditions. Crucial for the design and interpretation of indirect-drive ICF experiments (or any laser-driven high-$Z$ plasma for that matter) is a comprehensive understanding of (i) the interaction of high-intensity laser light with the plasma and (ii) the radiative properties of complex, high-$Z$ ions embedded in a plasma out of equilibrium, a so-called non-local thermodynamic equilibrium (non-LTE) plasma \cite{Ralchenko2016}. Both of these aspects are integral components of large-scale ``multi-physics'' codes often used to simulate indirect-drive ICF experiments \cite{Zimmerman1975,Marinak1996,Marinak2001}. 
The predictive capabilities of these simulations are often constrained by the approximations (and limitations) of the various physical models implemented in the codes, such as the treatment of non-LTE atomic kinetics, radiation transport, laser light absorption, etc. Adequate benchmarking of each of these aspects, be it through experimental comparisons \cite{Foord2000,Heeter2007} or code comparison efforts \cite{Piron2017,Gaffney2018,Grabowski2020}, is crucial for ensuring that their coupling in a multi-physics simulation yields reliable results. 

A second, more industrial-oriented application of laser-driven HEDP research is the study and characterization of extreme ultraviolet (EUV) plasma light sources for next-generation nanolithography \cite{Bakshi2006,OSullivan2015,Versolato2019}. In EUV lithography (EUVL), laser-driven plasmas formed on pre-deformed tin ($ Z = 50 $) microdroplet targets \cite{Kurilovich2016,Liu2020} act as the radiation source in the patterning of nano-scale features on integrated circuits \cite{Fomenkov2017,Felix2020}. Under optimum experimental conditions, EUV emission from such plasmas comprises an intense, narrowband feature (full width at half maximum $ \approx 0.6 $ nm \cite{vandeKerkhof2020}) centered near a wavelength of 13.5 nm. Importantly, only a small portion of this emission (13.5 $ \pm $ 0.135 nm $ - $ the so-called ``in-band'' region where molbydenum/silicon multilayer mirrors exhibit high reflectance \cite{Bajt2002}) is used to pattern features. To-date, plasma modeling has played an important role in guiding EUV light source development efforts. Prominent examples of this include the elucidation of EUV generation in laser-driven tin plasmas \cite{White2005,Poirier2006,Novikov2007,Koike2007,Sasaki2007,Sasaki2010,Torretti2020}, the characterization of tin-plasma properties \cite{MacFarlane2005,White2008,Nishihara2008,Basko2015,Basko2016,Su2017,Hemminga2021} as well as the identification of experimental conditions (laser parameters, tin target structures, etc.) which optimise the working conditions of the light source \cite{Cummings2005,Sizyuk2006,Sunahara2008,Purvis2016,Hassanein2013,Masnavi2019}. Looking to the future, the need to develop ever more powerful EUV sources (beyond the current 250 W of in-band EUV power \cite{Mayer2021}) will require (i) new, fundamental insights on EUV generation on complex tin targets \cite{Purvis2016} and (ii) exploring the plasma physics implications of alternate drive-laser concepts, e.g., the use of a Thulium-based 2-$\mu$m-wavelength drive laser \cite{Behnke2021,Tamer21}. Plasma modeling will play a crucial role in both regards. 

The task of modeling an EUV light source, which entails simulating laser interaction with a mid-$Z$, strongly radiating plasma, has many aspects in common with multi-physics ICF simulations. For one, EUV emission from laser-driven tin plasmas is determined by the coupling of non-LTE atomic kinetics and radiation transport through the plasma. While many codes have been developed for simulating EUV sources, they typically differ in their treatment of the underlying physical processes, e.g., the radiative properties of the plasma, equation-of-state (EOS) properties and the absorption and refraction of laser light. A multitude of factors can therefore contribute to deviations between codes. A well-known example of this is quantifying the so-called ``conversion efficiency'' of the light source (the ratio of in-band EUV energy emitted in the 2$\pi$ hemisphere back towards the laser to the laser energy \cite{Schupp2019}), which exhibits a heightened sensitivity to the accuracy of the radiative data (line positions and intensities) used in the simulations. Extensive benchmarking of the various physics components entering such codes is essential for developing predictive plasma modeling toolkits. While code-to-code benchmarking has been active in the ICF community for several years \cite{Gaffney2018,Grabowski2020,Lee1997,Rubiano2007,Fontes2009}, no such platform exists for the EUV light source plasma community.

In this paper, we summarize the findings of the 1$^{st}$ EUV Light Sources Code Comparison Workshop, which was held at the 2020 Source Workshop \cite{EUVlitho} (jointly organised by EUV Litho, Inc., the Paul Scherrer Institute and ETH Z{\"u}rich). In lieu of the ongoing coronavirus pandemic, the workshop was held online on 3$^{rd}$ November 2020. The goal of this workshop was to provide a platform for the EUV light source plasma modeling community to test and benchmark their codes using \textit{well-defined} test problems. In this first edition of the workshop, two aspects of EUV light source plasma modeling were addressed. The first case study investigated the atomic kinetics and radiative properties of tin plasmas under EUV-generating conditions. The second case study examined laser absorption in a one-dimensional, fully-ionized hydrogen plasma. The present paper is structured as follows: In Section 2, we describe the workshop structure and organization. This is followed by an overview of the results and key learnings of the (i) atomic kinetics case study (Section 3) and (ii) the laser absorption case study (Section 4). In Section 5 we outline plans for future code comparison activities, and the paper is concluded in Section 6.

\section{Workshop organization and structure}

The organization and structure of the 1$^{st}$ EUV Light Sources Code Comparison Workshop was modelled on the series of highly successful non-LTE code comparison workshops \cite{Lee1997,Rubiano2007,Fontes2009,Hansen2020}. Approximately four months prior to the workshop, the authors of this paper gathered to define a series of test problems to be investigated at a code comparison session to be held at the 2020 Source Workshop \cite{EUVlitho}. The problems were shared with prospective participants in August 2020. After a series of discussions, two independent case studies were defined. The first of these focused on the atomic kinetics and radiative properties (opacities, emissivities) of tin plasmas in EUV source-relevant conditions. This problem closely resembles a ``standard'' case study addressed at the non-LTE code comparison workshops. The second case study investigated laser absorption in a static, fully ionized, one-dimensional hydrogen plasma. While these plasma conditions do not represent those found in an EUV light source plasma, this problem served as a first, basic test of laser absorption routines. Moreover, it will provide a baseline upon which additional layers of complexity will be introduced, e.g., non-LTE radiation transfer effects, hydrodynamic motion, etc. Contributors were required to submit their results to the workshop committee one week prior to the workshop in a format similar to that used for the non-LTE code comparison workshops.

The first part of the workshop was devoted to code presentations, where the authors of the submissions described their codes and gave a brief overview of their results. 10 submissions were received for the atomic kinetics case study and 4 submissions were received for the laser absorption problem. A list of participating codes and contributors is provided in Table \ref{table1}. 
The final two presentations of the session were devoted to summarizing the results of the two case studies. In this paper, we will give an overview of the key findings of the workshop and will outline plans for future code comparison activities on the topic of EUV light source plasmas. 
As is standard practice in the proceedings of the non-LTE code comparison workshops, 
the published results are completely anonymous.

\begin{sidewaystable}
\begin{center}
\begin{tabular}{|c c c|} 
 \hline
 Code & Contributors & Institution (Country) \\ [0.5ex] 
 \hline
 \textit{Case study 1: Atomic kinetics} & & \\
 JATOM \cite{Sasaki2013} & A. Sasaki, K. Nishihara, A. Sunahara & KPSI, ILE, CMUXE (Japan, USA) \\ 
 ATOMIC \cite{Magee2004,Hakel2006} & J. Colgan & LANL (USA) \\
 THERMOS \cite{Nikiforov2006,Vichev2019} & I. Yu. Vichev, A. D. Solomyannaya, A. S. Grushin, D. A. Kim & KIAM (Russia) \\
 PrismSPECT \cite{MacFarlane2003} & I. E. Golovkin & Prism Comp. (USA) \\
 Cretin \cite{Scott2001} & H. A. Scott & LLNL (USA) \\ [1ex] 
 SEMILLAC \cite{Frank2013,Frank2014} & Y. Frank & L2X,LLNL (Israel,USA) \\ [1ex] 
 \textit{Case study 2: Laser absorption} & & \\
 STAR2D \cite{Sunahara2012,Sunahara2013} & A. Sunahara, K. Nishihara, A. Sasaki & CMUXE, ILE, KPSI (USA, Japan) \\ [1ex]
 RALEF-2D \cite{Basko2010,Basko2012} & M. M. Basko & KIAM  (Russia) \\ [1ex]
 HELIOS \cite{MacFarlane2006} & I. E. Golovkin & Prism Comp. (USA) \\ [1ex]
 Cretin \cite{Scott2001} & H. A. Scott & LLNL (USA) \\ [1ex]  
 \hline
\end{tabular}
\caption{List of codes and contributors.}
\label{table1}
\end{center}
\end{sidewaystable}

\section{Case study 1: Atomic kinetics of tin plasmas}

The objective of this case study was to investigate the atomic kinetics and radiative properties of zero-dimensional (``optically thin'') tin plasmas at EUV light source plasma conditions. 25 test cases were defined and specified according to the (i) electron density of the plasma $ n_{e} $ and (ii) electron temperature $ T_{e} $ (see Table \ref{table2} for a complete list of plasma conditions). The plasmas were considered quasineutral, i.e., $ n_{e} = Z^{*}n_{i} $ where $ Z^{*}$ is the average charge state of the plasma and $ n_{i} $ is the ion density. The ion temperature was specified to be equal to the electron temperature. The test cases do not consider any external radiation field-driven effects, i.e., the radiation temperature $ T_{r} = 0 $ eV for all test cases. The two ``bounding'' density cases $ n_{e} = 10^{19} $ and $ 10^{21}$ cm$^{-3}$ are of particular relevance for EUV light source plasmas as they correspond to the critical electron densities (the electron densities beyond which laser light does not propagate in the plasma) for CO$_{2}$ ($ \lambda_{\mathrm{laser}} = 10.6 $ $\mu$m) and Nd:YAG laser light ($ \lambda_{\mathrm{laser}} = 1.064 $ $\mu$m), respectively.  For each test case, participants were asked to compute:

\begin{itemize}
    \item The charge state distribution (CSD) and average charge state $ Z^{*}$ of the plasma.
    \item The emission ($ \eta_{\lambda}$) and absorption ($ \alpha_{\lambda} $) coefficients.
    \item The spectral purity (SP) of the plasma, defined here as the ratio of the emissivity in the in-band region of molybdenum/silicon multilayer mirrors (13.5 $ \pm $ 0.135 nm) to the emissivity in the 5 $ - $ 20 nm region, i.e., $ \mathrm{SP} $ $ (\%) = 100 \times \int_{13.365}^{13.635}\eta_{\lambda}d\lambda/\int_{5}^{20}\eta_{\lambda}d\lambda $.
    \item The specific internal energy (SIE), defined as SIE = $ \sum_{j}E_{j}n_{j} $ where the sum runs over all states $ j $ (levels, configurations, etc.) having energy $ E_{j} $ and population density (simply ``population'' in the following) $ n_{j} $.
    \item Contributions to the radiative power losses (RPL): bound-bound, bound-free and free-free transitions.
\end{itemize}

\begin{table}
\centering
\begin{tabular}{|c c|} 
 \hline
 $n_{e}$ (cm$^{-3}$) & $ T_{e} $ (eV) \\ [0.5ex] 
 \hline
 10$^{19}$ & 10, 15, 20, 25, 30, 35, 40 \\
 10$^{20}$, 10$^{21}$ & 20, 25, 30, 35, 40, 45, 50, 55, 60 \\
 \hline
\end{tabular}
\caption{Plasma conditions for the atomic kinetics case study.}
\label{table2}
\end{table}

\noindent In the following, we divide our discussions into three cases according to the electron density: $ n_{e} = 10^{19} $ cm$^{-3}$ (case 1), $ 10^{20}$ cm$^{-3}$ (case 2) and $ 10^{21}$ cm$^{-3}$ (case 3).

\subsection{Case 1: $ n_{e} = 10^{19} $ cm$^{-3}$}

\begin{figure}
    \centering
\includegraphics[scale=0.8]{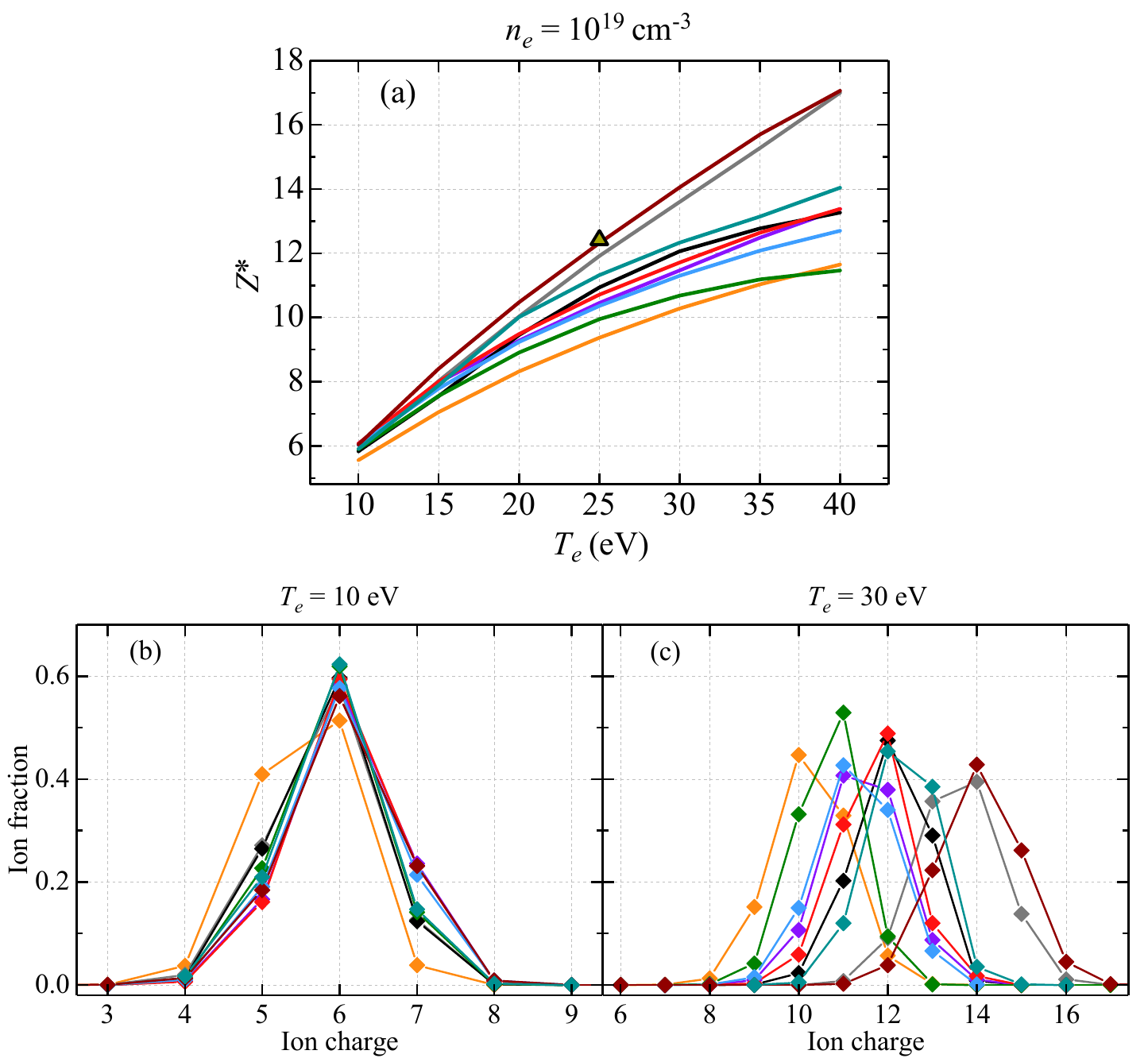}
    \caption{(a) Average charge state $ Z^{*} $ of the plasma as a function of $ T_{e} $ for $ n_{e} = 10^{19}$ cm$^{-3}$. Charge state distributions of the (b) $ T_{e} $ = 10 eV and (c) $ T_{e} $ = 30 eV plasmas. The gray, dark red and gold triangle submissions were calculated under the assumption of LTE conditions.}
    \label{fig:Zbar_CSD_10_19}
\end{figure}

In Fig.\,\ref{fig:Zbar_CSD_10_19} (a) we present computations of the average charge state $ Z^{*} $ of tin plasmas having $ n_{e} = 10^{19} $ cm$^{-3}$ and $ T_{e} = 10 - 40 $ eV. This electron density is of particular relevance for industrial applications as it corresponds to the critical electron density for CO$_{2}$ lasers which are currently used to drive the EUV-emitting plasma \cite{vandeKerkhof2020,Torretti2019}. The charge state distributions (CSDs) of the $ T_{e} = 10 $ and 30 eV plasmas are shown in Fig.\,\ref{fig:Zbar_CSD_10_19} (b) and (c), respectively.

First, for the lowest temperature case ($ T_{e} = 10 $ eV), we see that all codes predict a similar value of $ Z^{*} \approx 6 $. This convergence in $ Z^{*} $ is reflected in Fig.\,\ref{fig:Zbar_CSD_10_19} (b), where only minor differences in the shapes of the CSDs are observed between most codes. The plasma conditions (high-density, low-temperature) place it close to LTE conditions, and thus agreement between the codes in this region is expected as only the atomic structure matters in determining the ionization balance in LTE. With increasing electron temperature, $ Z^{*} $ increases alongside an increased spread in the various code predictions. This is most apparent for temperatures above 30 eV, where predictions of $ Z^{*} $ span six charge states. As the electron temperature increases, the plasma moves away from LTE conditions and enters the non-LTE regime, a regime known for producing disagreements among collisional-radiative codes \cite{Chung2013}. It should be noted that the calculations shown in gray, dark red and the single gold triangle were performed under the assumption of LTE conditions. 
These LTE calculations, which are equivalent to performing Planckian radiation field-driven non-LTE calculations with $ T_{r} = T_{e} $, naturally overestimate the plasma ionicity as we have specified $ T_{r} = 0 $ eV in the current problem. The spread in $ Z^{*} $ predictions likely originate from differences in the dielectronic recombination rates, a supposition suggested by investigations undertaken in the non-LTE code comparison workshops \cite{Chung2013}. Atomic model completeness, most notably extensive consideration of autoionizing channels, is an important aspect of ionization balance calculations \cite{Chung2013,Hansen_CR_modeling} which will be addressed in a future iteration of this workshop.

\begin{figure}
    \centering
\includegraphics[scale=0.5]{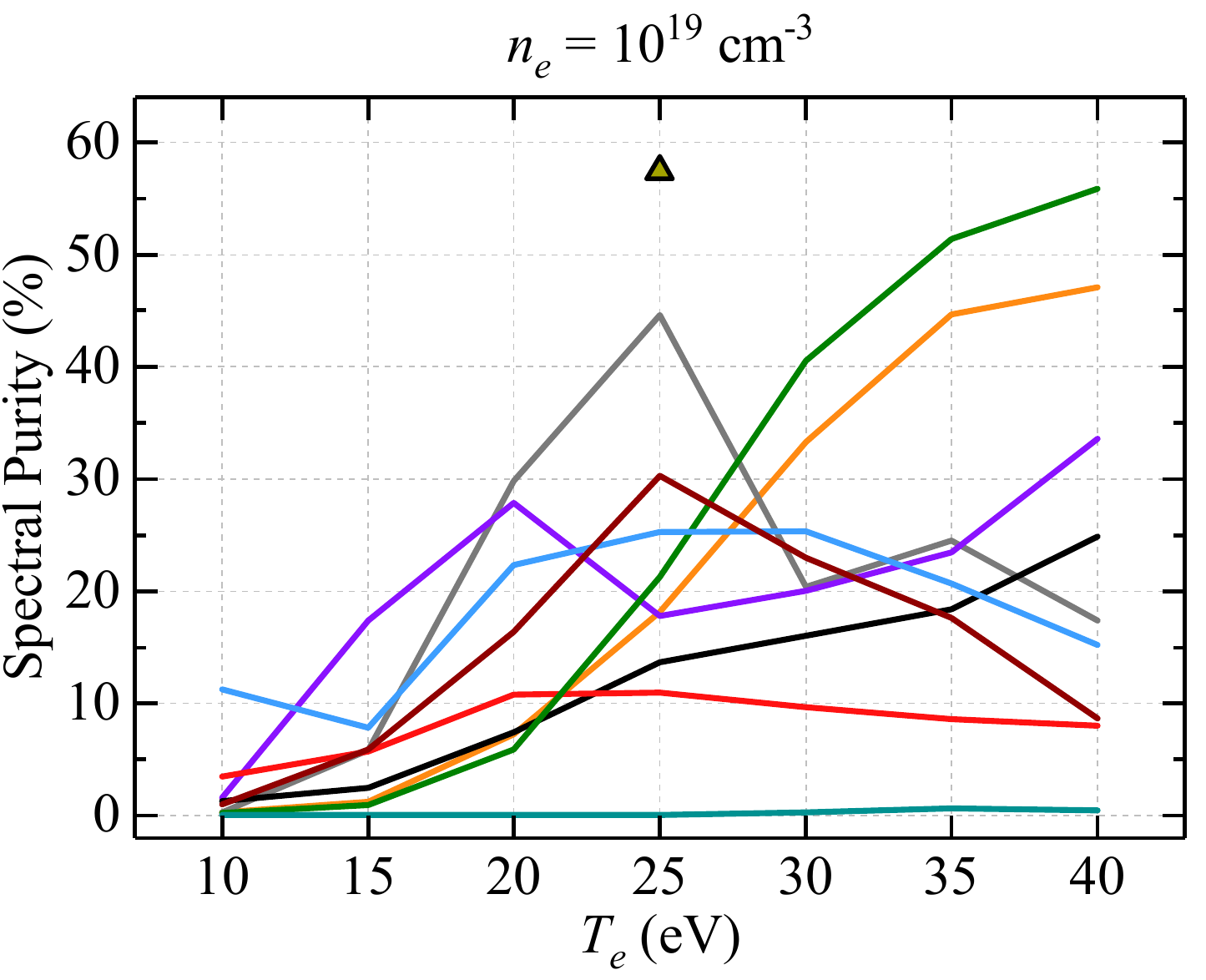}
    \caption{Spectral purity as a function of $ T_{e} $ for $ n_{e} = 10^{19}$ cm$^{-3}$. The gray, dark red and gold triangle submissions were calculated under the assumption of LTE conditions.}
    \label{fig:SP_10_19}
\end{figure}

\begin{figure}
    \centering
\includegraphics[scale=0.55]{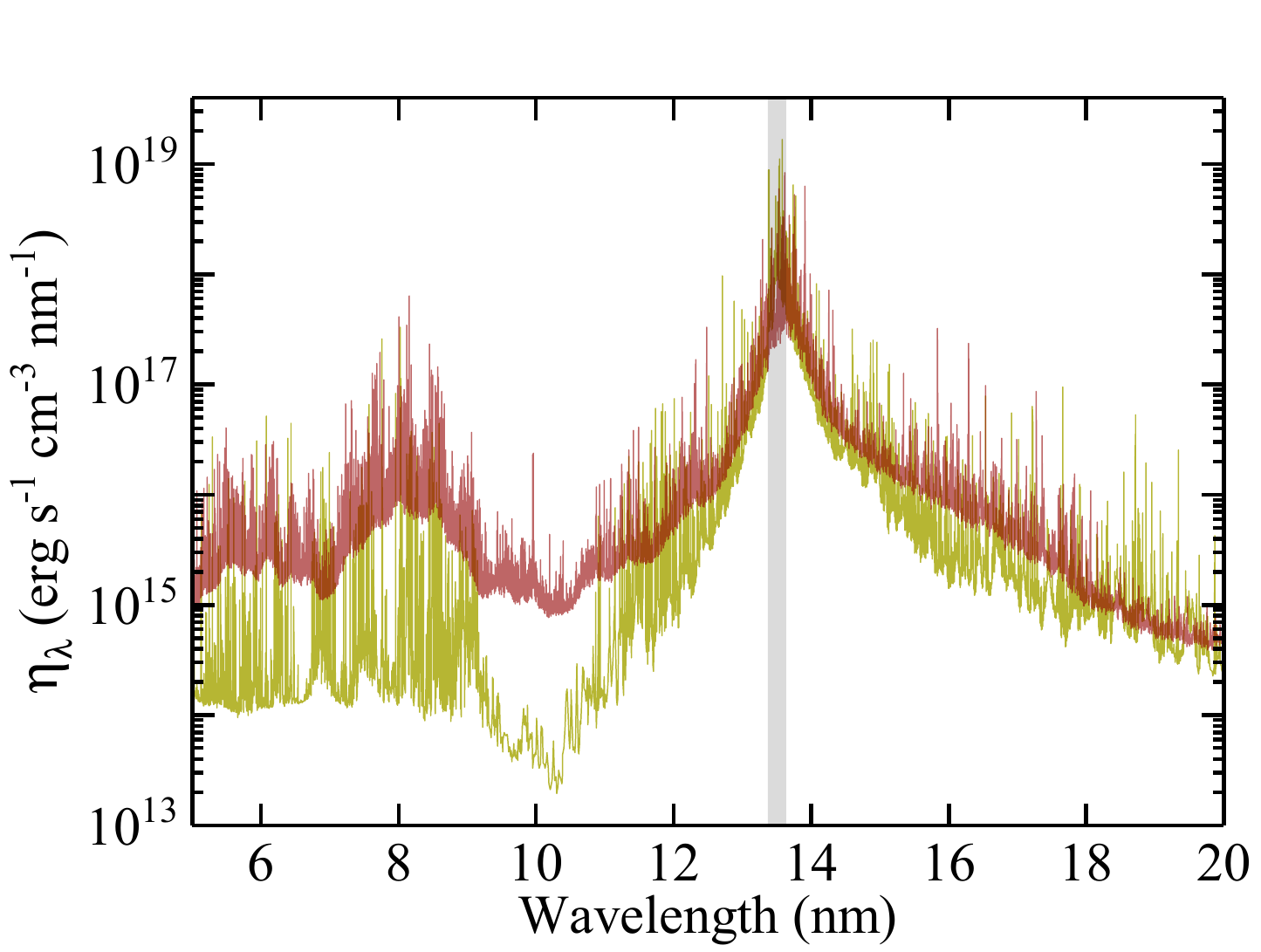}
    \caption{Emissivity of the $ n_{e} = 10^{19}$ cm$^{-3}$, $ T_{e} = 25 $ eV plasma as calculated by the dark red and gold triangle LTE submissions. The gray shaded area represents the $ 13.5 \pm 0.135 $ nm in-band region.}
    \label{fig:EUV_spectra_LTE_10_19}
\end{figure}

In Fig.\,\ref{fig:SP_10_19}, we present calculations of the spectral purity (SP) of these plasmas. Unlike the $ Z^{*} $ calculations, no general trend exists among the codes. First, we note that the LTE submissions (gold triangle, gray and dark red curves) predict peaks in the SP for $ T_{e} = 25 $ eV. This plasma condition generates an average charge state $ Z^{*} \equiv Z^{*}_{\mathrm{SP}_{\mathrm{peak}}} \approx 12 $ for all three LTE submissions (see Fig.\,\ref{fig:Zbar_CSD_10_19} (a)). While the codes agree on $ Z^{*}_{\mathrm{SP_{\mathrm{peak}}}} $, the spread in SP is significant: 30$\%$ (dark red), 45$\%$ (gray) and 60$\%$ (gold triangle). In Fig.\,\ref{fig:EUV_spectra_LTE_10_19}, we plot the emissivity of the $ T_{e} = 25 $ eV plasma as calculated by the dark red and gold triangle submissions. The atomic model associated with the dark red spectrum includes an extensive number of multiply excited states in the underlying atomic structures, and transitions from such states leads to significantly more emission in the $ 5 - 13 $ nm region than the gold triangle submission. The spectrum shown in gold exhibits (i) more intense in-band emission and (ii) less ``out-of-band'' emission than the dark red spectrum. Both of these factors contribute to the higher spectral purity for the gold triangle submission. 
Returning to Fig.\,\ref{fig:SP_10_19}, we note that the green and orange non-LTE submissions exhibit a steep rise in SP for $ T_{e} > 20 $ eV, where the ``global'' SP maximum is most likely achieved for $ T_{e} > 40 $ eV. The blue and bright red submissions, on the contrary, predict a slowly varying SP with temperature, both of which peak for $ Z^{*}_{\mathrm{SP}_{\mathrm{peak}}} \approx 11 $ (SP = $ 25 \% $ and $ 10 \% $, respectively). Finally, we note the local SP maximum for the purple curve at $ T_{e} = 20 $ eV is non-physical, arising from a misplaced shifting of spectral lines.

\begin{figure}
    \centering
\includegraphics[scale=0.55]{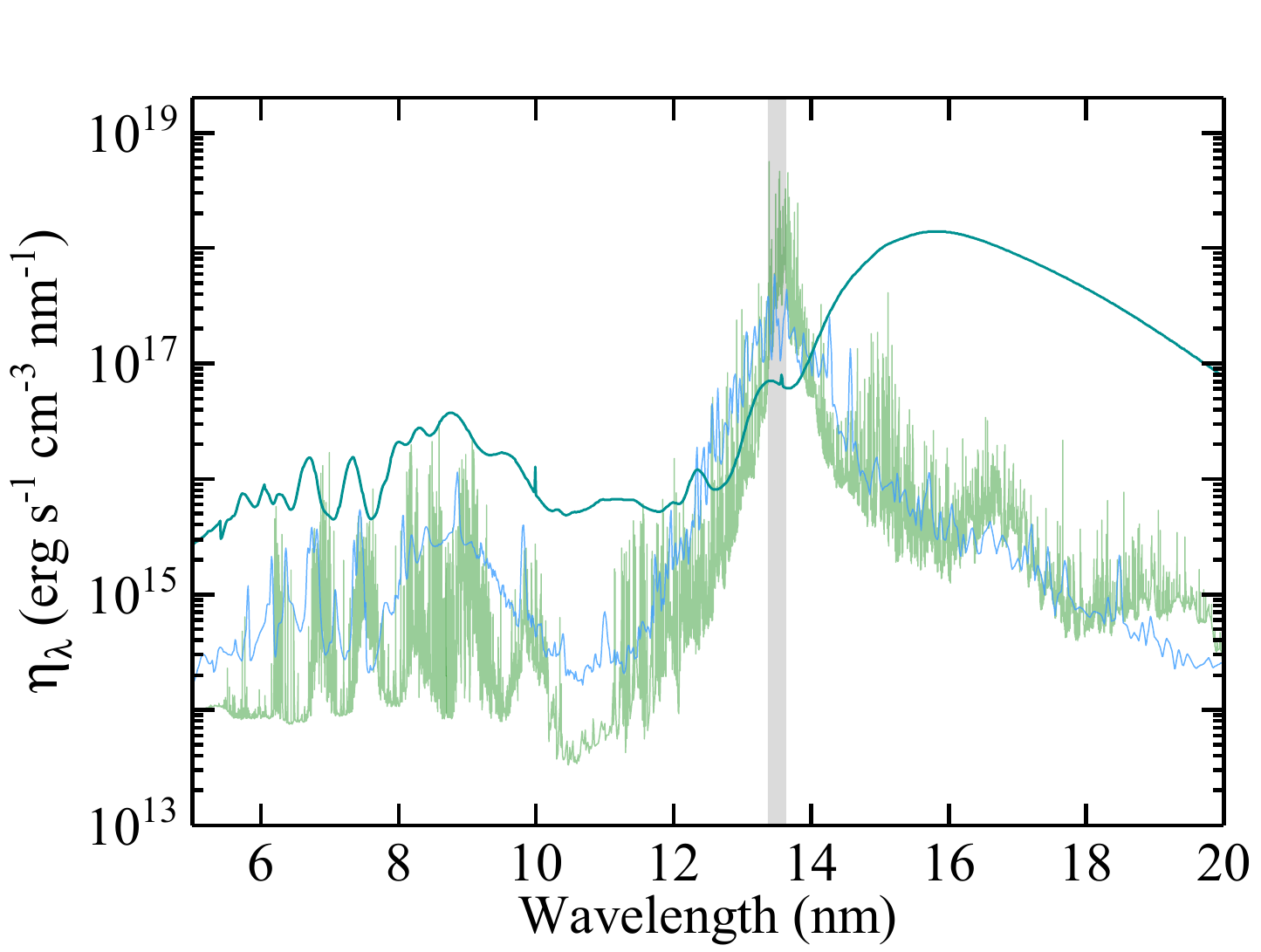}
    \caption{Plasma emissivities as calculated by the blue, green and dark cyan non-LTE submissions for plasmas having $ Z^{*} \approx 11 $. The gray shaded area represents the $ 13.5 \pm 0.135 $ nm in-band region.}
    \label{fig:EUV_spectra_nLTE_10_19}
\end{figure}

\begin{figure}
    \centering
\includegraphics[scale=1]{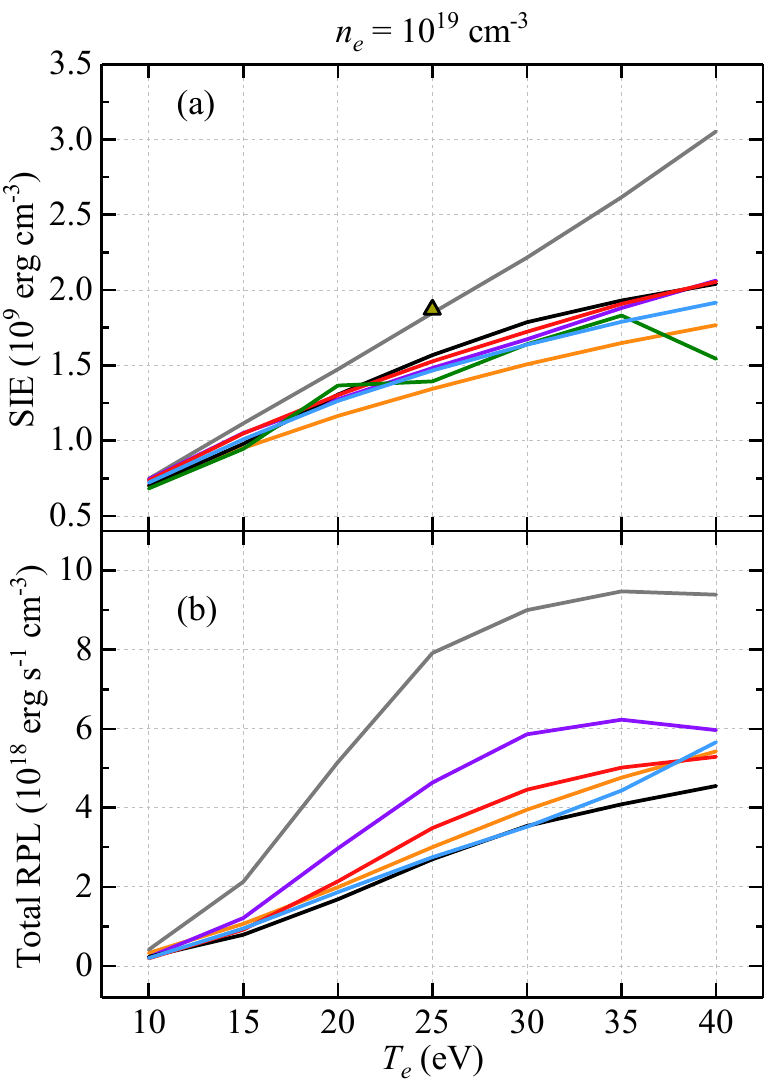}
    \caption{(a) Specific internal energy (SIE) and (b) total radiative power losses (RPL) as a function of $ T_{e} $ for $ n_{e} = 10^{19}$ cm$^{-3}$. The gray and gold triangle submissions were calculated under the assumption of LTE conditions.}
    \label{fig:eint_ploss_10_19}
\end{figure}

The absence of a common trend among the non-LTE submissions shown in Fig.\,\ref{fig:SP_10_19} is perhaps not too surprising considering (i) the substantial spread in $ Z^{*} $ predictions for $ T_{e} > 15 $ eV (Fig.\,\ref{fig:Zbar_CSD_10_19} (a)) (ii) differences in the underlying atomic structures of the tin charge states and (iii) the rather narrow bandwidth (0.27 nm) associated with the in-band region. This is exemplified in Fig.\,\ref{fig:EUV_spectra_nLTE_10_19} where we plot the emissivity of plasmas having $ Z^{*} \approx 11 $ for the green, blue and dark cyan non-LTE submissions ($ T_{e} = 35, 30 $ and $ 25 $ eV plasmas, respectively). The dark cyan emissivity is clearly very different to that of the green and blue emissivities, the latter of which predicts substantially less in-band emission that the spectrum shown in green. It is well known that the atomic structures of Sn$^{10+}$ $ - $ Sn$^{14+}$ ions are subject to strong configuration effects, and this makes their accurate calculation notoriously difficult \cite{Koike2007,Dortan2007}. Extensive benchmarking of these level structures with charge state-specific experimental spectra \cite{DArcy2009a, Ohashi2010, DArcy2011,Scheers2020,Bouza2020} and/or highly accurate \textit{ab initio} atomic structure calculations \cite{Windberger2016,Scheers2020,Filin2021} is therefore required. Experimental measurements of emission from multiply excited states are also highly desired given their dominant contribution to EUV emission from laser-driven, mid-$Z$ plasmas \cite{Sasaki2004,Torretti2020,Sheil2021}. 

Finally, in Fig.\,\ref{fig:eint_ploss_10_19} we plot the (a) specific internal energy (SIE) and (b) total radiative power losses (RPL) for the $ n_{e} = 10^{19} $ cm$^{-3}$ plasma as a function of $ T_{e} $. The LTE submissions (gray curve, gold triangle) predict a higher SIE than the non-LTE submissions. For the green submission, we notice a reduction in the SIE in going from $ T_{e} = 35 $ to 40 eV. This behaviour is rather unusual given the minor change in $ Z^{*}$ between these two cases, where $ Z^{*}(T_{e} = 35$ eV$) $ $ \approx 11.2 $ and $ Z^{*}(T_{e} = 40 $ eV$) $ $ \approx 11.5 $. Generally speaking, there is good agreement between the codes for the SIE. Moving to the total RPL, we see that the gray, purple, bright red and black submissions all exhibit a similar ``s-like'' shape with a near-plateau in the total RPL for $ T_{e} > 30 $ eV. 

\subsection{Case 2: $ n_{e} = 10^{20}$ cm$^{-3}$}

\begin{figure}
    \centering
\includegraphics[scale=1]{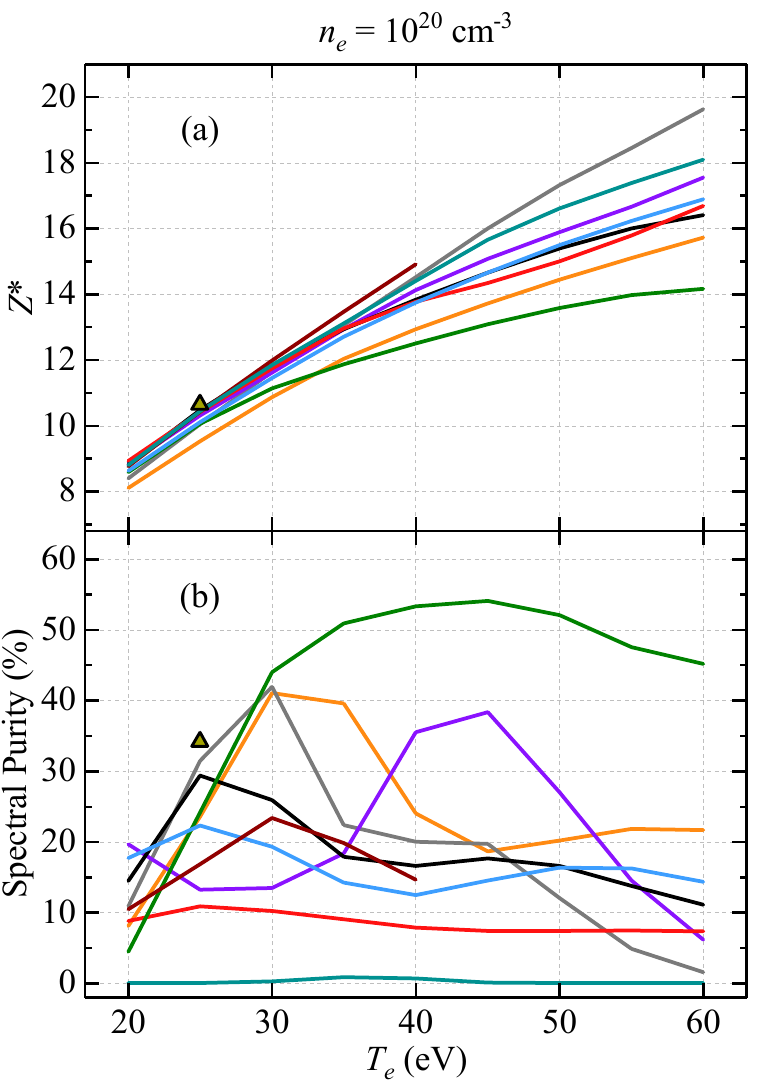}
    \caption{(a) Average charge state $ Z^{*} $ and (b) spectral purity as a function of $ T_{e} $ for $ n_{e} = 10^{20}$ cm$^{-3}$. The gray, dark red and gold triangle submissions were calculated under the assumption of LTE conditions.}
    \label{fig:Zbar_SP_20}
\end{figure}

In Fig.\,\ref{fig:Zbar_SP_20} (a) we plot $ Z^{*} $ as a function of $ T_{e} $ for plasmas with $ n_{e} = 10^{20}$ cm$^{-3} $. As in the $ n_{e} = 10^{19}$ cm$^{-3}$ case, best agreement between $ Z^{*}$ predictions is achieved for the low temperature ($ T_{e} = 20/25 $ eV) cases. 
For the highest temperature case ($ T_{e} = 60 $ eV), we see that the majority of the non-LTE submissions are clustered in the $ Z^{*} = 16 - 18 $ range. The spectral purities calculated by these codes are shown in Fig.\,\ref{fig:Zbar_SP_20} (b). Unlike the $ n_{e} = 10^{19}$ cm$^{-3} $ case, we can identify ``global'' peaks in the spectral purity for the black, purple, green and orange codes. The average charge state for which the spectral purity peaks, $ Z^{*}_{\mathrm{SP}_{\mathrm{peak}}} $, is found to be $ Z^{*}_{\mathrm{SP}_{\mathrm{peak}}} \approx 10 $ (blue submissions), 10.5 (red and black submissions), 11 (orange submission), 12 (gray and dark red LTE submissions), 13 (green submission) and 15 (purple submission). While the orange curve also exhibits a large SP for $ Z^{*} \approx 12 $, the spread in $ Z^{*}_{\mathrm{SP}_{\mathrm{peak}}} $ (and SP values) among the various codes is substantial. These observations reinforce the need for extensive benchmarking of atomic spectra calculations and population kinetics models.

\begin{figure}
    \centering
\includegraphics[scale=1]{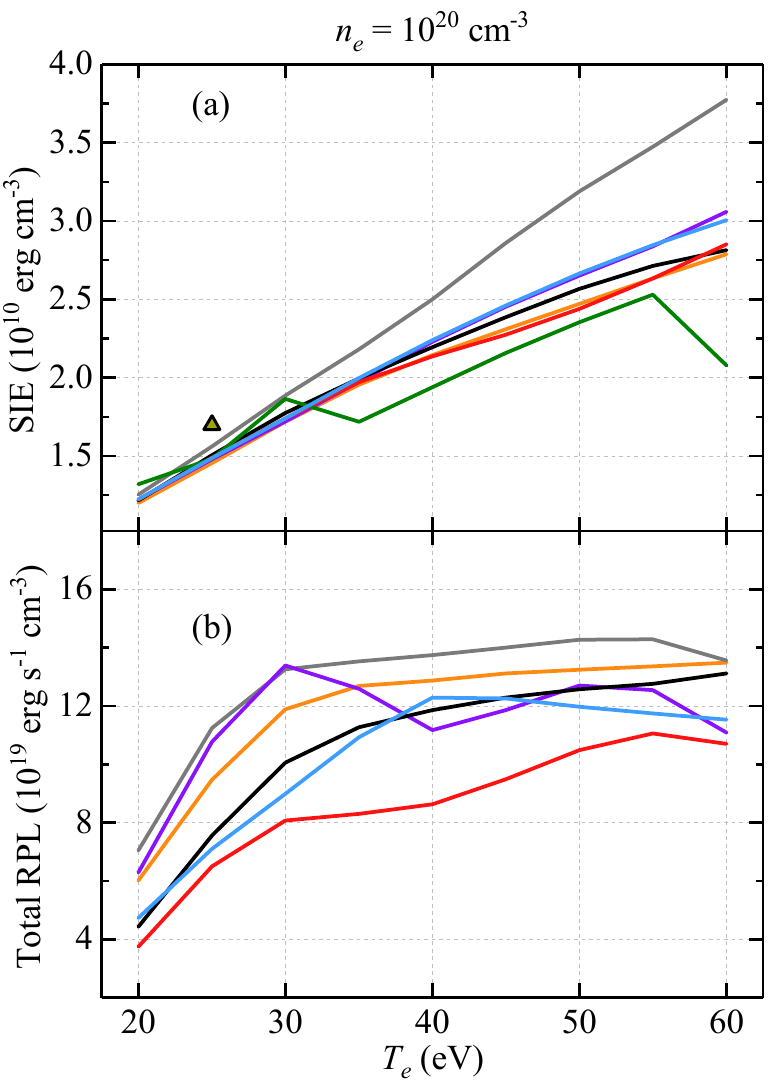}
    \caption{(a) Specific internal energy (SIE) and (b) total radiative power losses (RPL) as a function of $ T_{e} $ for $ n_{e} = 10^{20}$ cm$^{-3}$. The gray and gold triangle submissions were calculated under the assumption of LTE conditions.}
    \label{fig:eint_ploss_20}
\end{figure}

The SIE and total RPL as a function of $ T_{e} $ for $ n_{e} = 10^{20}$ cm$^{-3} $ are shown in Fig.\,\ref{fig:eint_ploss_20} (a) and (b), respectively. First, the order-of-magnitude increase in $ n_{e} $ ($ 10^{19} \rightarrow 10^{20} $ cm$^{-3}$) gives rise to an order-of-magnitude increase in both the SIE and total RPL. Examining Fig.\,\ref{fig:eint_ploss_20} (a), we note the existence of two ``dips'' in the green SIE curve, one at $ T_{e} = 35 $ eV ($Z^{*} \approx 11.9 $) and the other at $ T_{e} = 60 $ eV ($Z^{*} \approx 14.2 $). While the exact cause of this behaviour is unknown, it may be due to irregularities in the underlying atomic structures of certain ions. The remaining non-LTE submissions (black, orange, bright red, purple and blue curves) are in very good agreement with each other across the studied temperature range. The total RPL curves (excluding that of the bright red submission) exhibit a similar trend: an initial steep rise at low temperatures followed by a near-plateauing of the total RPL. We note that the purple submission exhibits a peak in the total RPL at $ T_{e} = 30 $ ($ Z^{*} \approx 11.5$). 

\subsection{Case 3: $ n_{e} = 10^{21}$ cm$^{-3}$}

\begin{figure}
    \centering
\includegraphics[scale=1]{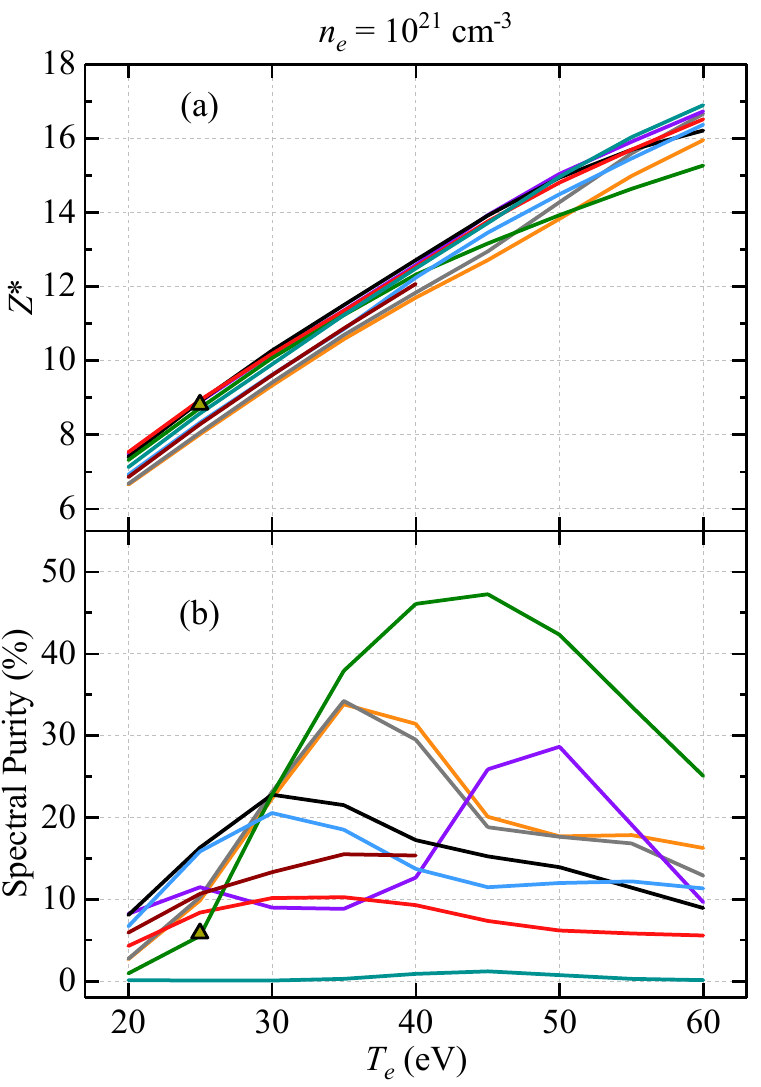}
    \caption{(a) Average charge state $ Z^{*} $ and (b) spectral purity as a function of $ T_{e} $ for $ n_{e} = 10^{21}$ cm$^{-3}$. The gray, dark red and gold triangle submissions were calculated under the assumption of LTE conditions.}
    \label{fig:Zbar_SP_21}
\end{figure}

\begin{figure}
    \centering
\includegraphics[scale=0.55]{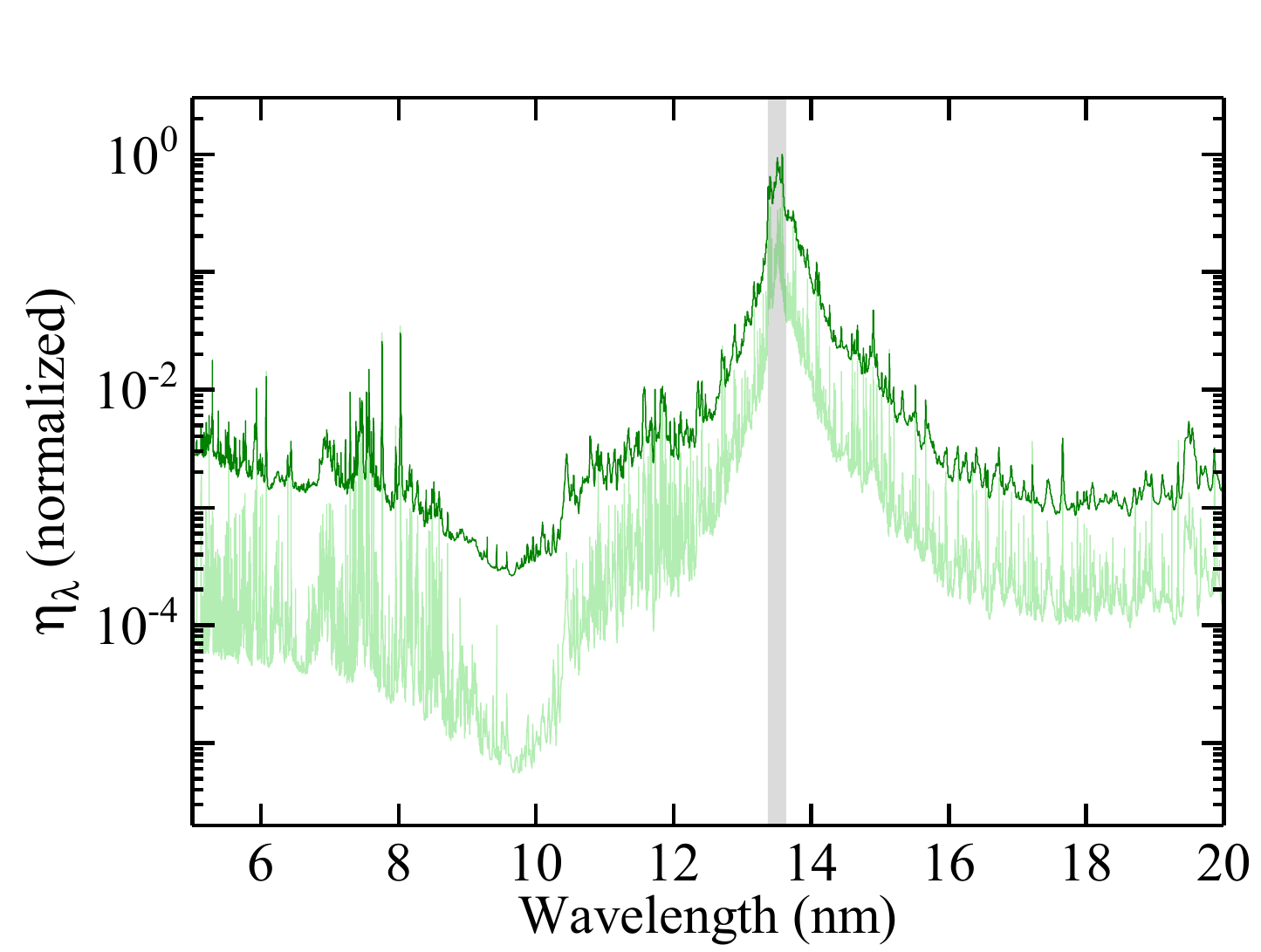}
    \caption{Normalized emissivities of the $ T_{e} = 45 $ eV, $ n_{e} = 10^{20}$ cm$^{-3}$ plasma (light green) and the $ T_{e} = 45 $ eV, $ n_{e} = 10^{21}$ cm$^{-3}$ (dark green) plasma. The gray shaded area represents the $ 13.5 \pm 0.135 $ nm in-band region.}
    \label{fig:EUV_spectra_nLTE_20_21}
\end{figure}

\begin{figure}
    \centering
\includegraphics[scale=1]{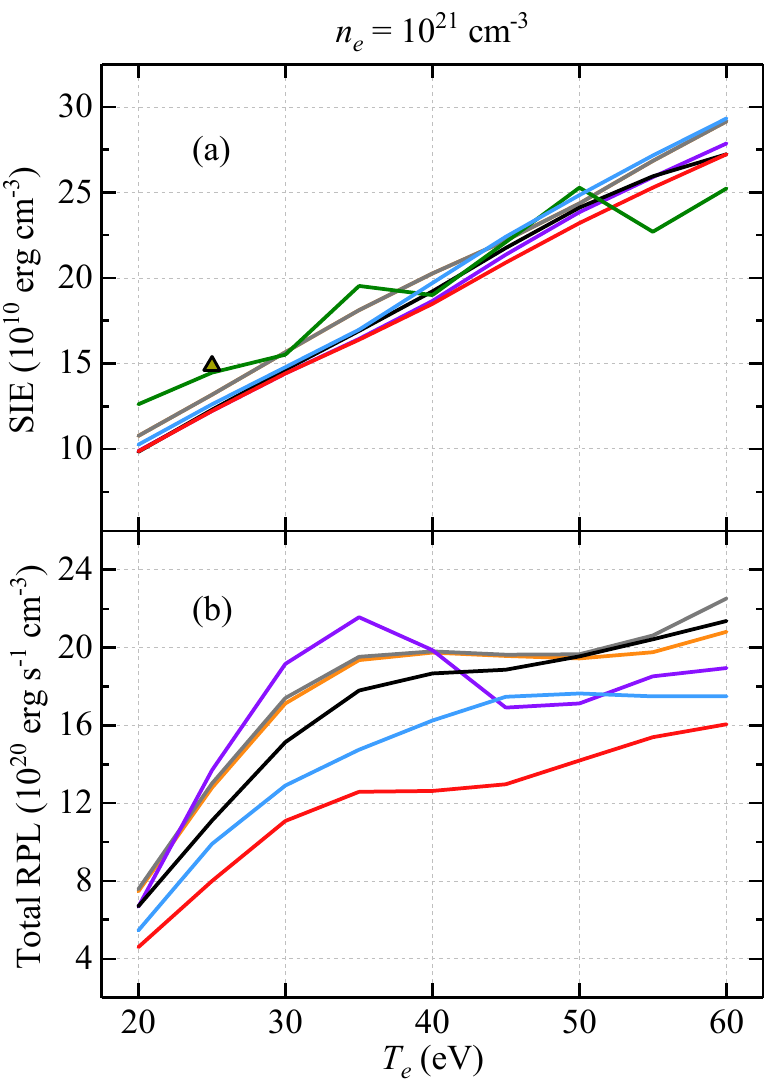}
    \caption{(a) Specific internal energy (SIE) and (b) total radiative power losses (RPL) as a function of $ T_{e} $ for $ n_{e} = 10^{21}$ cm$^{-3}$. The gray and gold triangle submissions were calculated under the assumption of LTE conditions.}
    \label{fig:eint_ploss_21}
\end{figure}

Finally, we discuss the high-density $ n_{e} = 10^{21} $ cm$^{-3}$ case. In line with case 2, we show in Fig.\,\ref{fig:Zbar_SP_21} (a) the average charge state $ Z^{*} $ and (b) spectral purity of the $ n_{e} = 10^{21} $ cm$^{-3}$ plasmas as a function of $ T_{e} $. The vast majority of the codes exhibit excellent agreement in $ Z^{*} $ in the studied $ T_{e} $ range. In fact, the plasma conditions considered here are close to LTE conditions, and this is reason for the good agreement amongst the codes. 
Examining Fig.\,\ref{fig:Zbar_SP_21} (b), we see that the SP curves all exhibit a similar trend to the $ n_{e} = 10^{20} $ cm$^{-3}$ case, however the maximum SP reached in the $ n_{e} = 10^{20} $ cm$^{-3}$ case is higher than that achieved in the higher-density $ n_{e} = 10^{21} $ cm$^{-3}$ case. To investigate this further, we show in Fig.\,\ref{fig:EUV_spectra_nLTE_20_21} the plasma emissivities as calculated by the green submission for the $ n_{e} = 10^{20} $ cm$^{-3}$, $ T_{e} = 45 $ eV case (shown in light green in Fig.\,\ref{fig:EUV_spectra_nLTE_20_21}) and the $ n_{e} = 10^{21} $ cm$^{-3}$, $ T_{e} = 45 $ eV case (shown in dark green in Fig.\,\ref{fig:EUV_spectra_nLTE_20_21}). Both of these plasmas exhibit $ Z^{*} \approx 13 $. It is clear that the $ n_{e} = 10^{21} $ cm$^{-3}$ spectrum exhibits more in- and out-of-band emission compared to the $ n_{e} = 10^{20} $ cm$^{-3}$ case. 
It is also interesting to note the two-order of magnitude difference in emissivities in the 5 $ - $ 10 nm range between the two plasma density cases. In Fig.\,\ref{fig:eint_ploss_21}, we plot the (a) SIE and (b) total RPL as a function of $ T_{e} $ for the $ n_{e} = 10^{21} $ cm$^{-3}$ case. In general, good agreement is found between the codes for the SIE. In terms of the total RPL, the submissions generally follow the same trend as the $ n_{e} = 10^{20} $ cm$^{-3}$ case. As in the $ n_{e} = 10^{20} $ cm$^{-3}$ case, the purple submission exhibits a peak in the total RPL for $ Z^{*} \approx 11.5$ ($ T_{e} = 35 $ eV).

\section{Case study 2: Laser absorption in a fully ionized hydrogen plasma}

We will now discuss the second case study addressed at the workshop, namely an investigation of laser absorption in a fully ionized hydrogen plasma. As mentioned in Section 2, this problem served as a first, basic test of laser absorption routines used in radiation-hydrodynamic simulations. As will be discussed in Section 5, we will introduce additional layers of complexity into this problem in future iterations of the workshop.

\begin{figure}
    \centering
\includegraphics[scale=1]{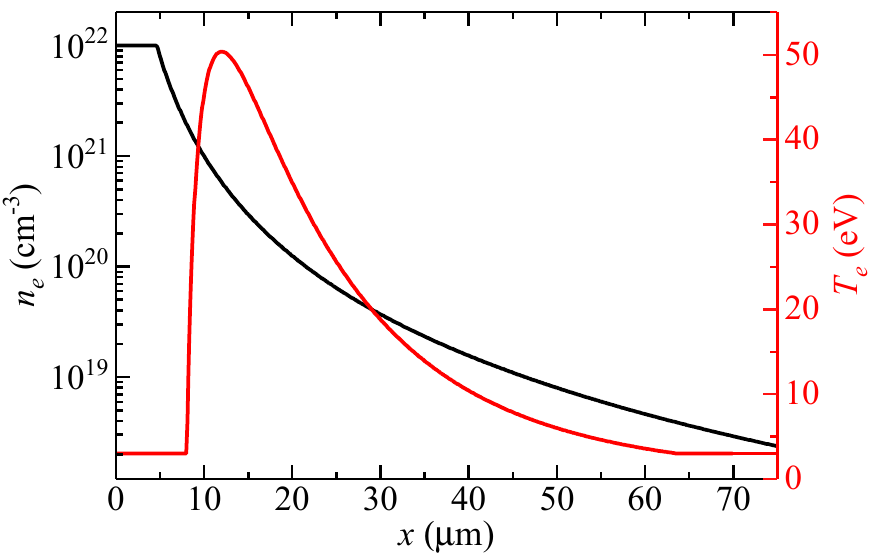}
    \caption{Profiles of the electron density $ n_{e} $ (black) and electron temperature $ T_{e} $ (red) for $ x \in [0, 75] $ $ \mu $m in the laser absorption case study. Note that the computational mesh extends to $ x = 300 $ $ \mu$m.}
    \label{fig:Electron_density_temperature}
\end{figure}

The problem setup is as follows: the plasma has a one-dimensional planar geometry with a computational domain defined over $ x \in [0, 300] $ $ \mu $m. The spatial mesh consists of 20\,000 equally spaced zones over this domain. The plasma parameters (temperature and density) are constant in time and are specified according to 

\begin{eqnarray}
    n_{e} = \mathrm{min}\left(10^{22},\frac{10^{24}}{x^{3}}\right) \mathrm{cm^{-3}} \\
    T_{e} = \mathrm{max}[3,93y\exp{(-y^{1/2})}]  \mathrm{eV}
\end{eqnarray}

\noindent where $ y = \mathrm{max}(0,x - 8) $. These profiles are illustrated in Fig.\,\ref{fig:Electron_density_temperature}. As the plasma consists of fully ionized hydrogen, the number density specifies both the electron and ion densities. The laser pulse is incident at $ x = 300 $ $ \mu $m at normal incidence, and laser absorption should be modelled using inverse bremsstrahlung. Two laser pulse wavelengths were considered in this study, namely $ \lambda_{\mathrm{laser}} = 1.064 $ $ \mu $m (Nd:YAG laser) and $ \lambda_{\mathrm{laser}} = 10.6 $ $ \mu $m (CO$_{2}$ laser). The incident laser intensity is $ 10^{11} $ W/cm$^{2}$ for both laser wavelength cases. The plasma is stationary (no hydrodynamic motion) and energy transfer processes such as radiation transport and thermal conduction are to be omitted. The requested quantities include the incident and reflected laser power densities, the deposited laser power, the laser absorption coefficient, and the (electron-ion) Coulomb logarithm.

\begin{figure}
    \centering
\includegraphics[scale=1]{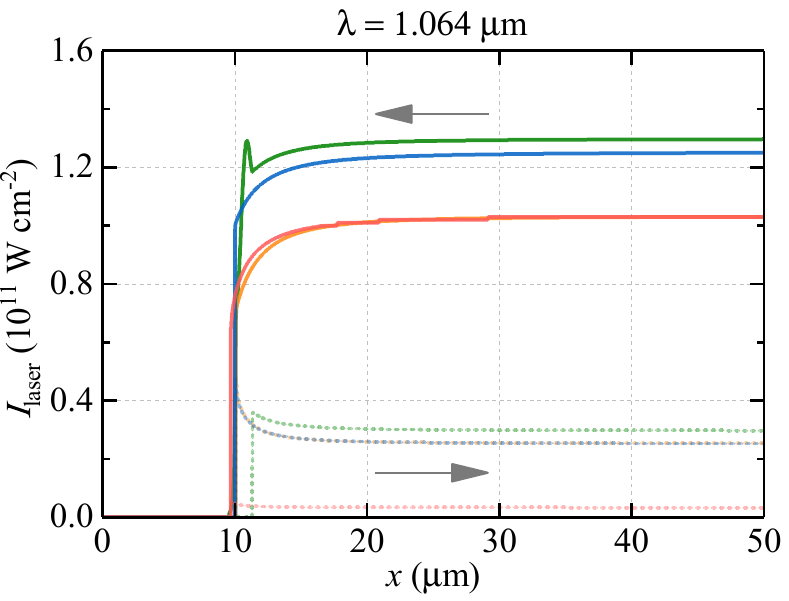}
    \caption{Laser power density $ I_{\mathrm{laser}} $ as a function of distance $ x $ for $ \lambda = 1.064 $ $ \mu $m. Solid lines correspond to the total laser power density. Dotted lines represent the power density of the reflected laser light. The gray arrows indicate the in-going and out-going laser light path.}
    \label{fig:Laser_Power_Density_1_micron}
\end{figure}

\begin{figure}
    \centering
\includegraphics[scale=1]{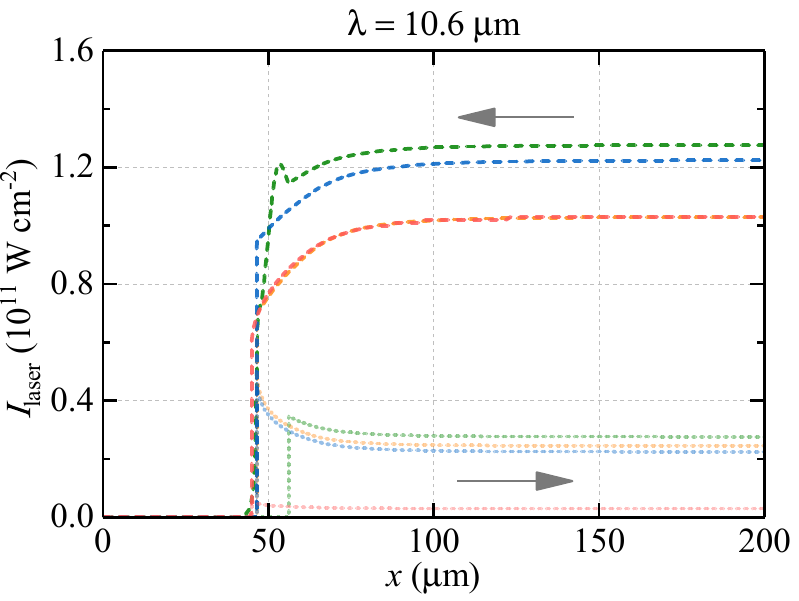}
    \caption{Laser power density $ I_{\mathrm{laser}} $ as a function of distance $ x $ for $ \lambda = 10.6 $ $ \mu $m. Dashed lines correspond to the total laser power density. Dotted lines represent the power density of the reflected laser light. The gray arrows indicate the in-going and out-going laser light path.}
    \label{fig:Laser_Power_Density_10_micron}
\end{figure}

\begin{figure}
    \centering
\includegraphics[scale=1]{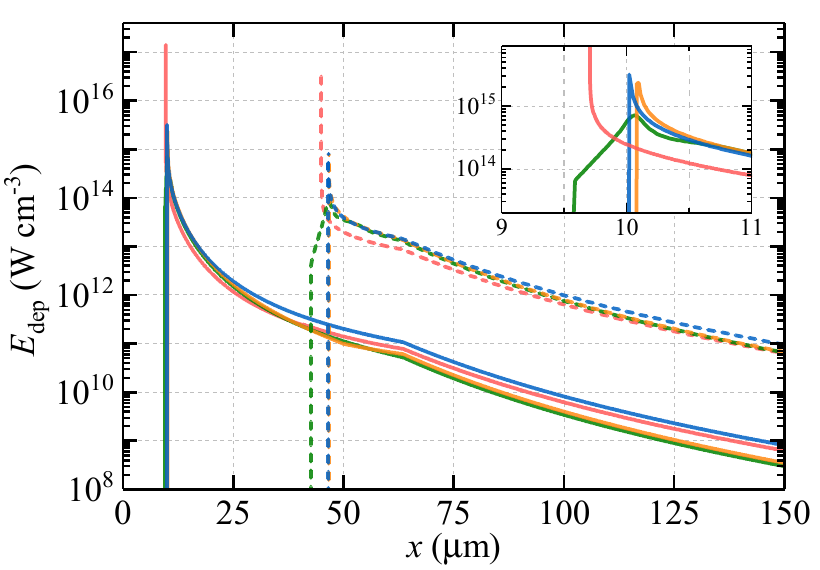}
    \caption{Laser energy deposition $ E_{\mathrm{dep}} $ as a function of distance $ x $. Solid lines corresponds to $ \lambda_{\mathrm{laser}} = 1.064 $ $ \mu $m and the dashed lines correspond to $ \lambda_{\mathrm{laser}} = 10.6 $ $ \mu$m. The inset figure provides a zoomed-in view near the critical density of the $ \lambda_{\mathrm{laser}} = 1.064 $ $ \mu $m case.}
    \label{fig:Laser_energy_deposition}
\end{figure}

In Fig.\,\ref{fig:Laser_Power_Density_1_micron}, we plot the laser power density $ I_{\mathrm{laser}} $ as a function of distance $ x $ for $ \lambda_{\mathrm{laser}} = 1.064 $ $ \mu $m. The dark solid lines indicate the total laser power density incident on the plasma (indicated by the arrow pointing towards $ x = 0 $ $\mu$m) and the light dotted lines represent the power density of the reflected laser light (indicated by the arrow pointing towards $ x = 300 $ $\mu$m). A similar plot for the $ \lambda_{\mathrm{laser}} = 10.6 $ $ \mu $m case is shown in Fig.\,\ref{fig:Laser_Power_Density_10_micron}. As expected, the laser light propagates up the critical electron density $ n_{\mathrm{crit}} \approx 10^{21}/\lambda^{2}_{\mathrm{laser}} $, which occurs at $ x \approx 10 $ $\mu$m for $ \lambda_{\mathrm{laser}} = 1.064 $ $ \mu$m and $ x \approx 45 $ $\mu$m for $ \lambda_{\mathrm{laser}} = 10.6 $ $ \mu$m. In Fig.\,\ref{fig:Laser_energy_deposition}, we plot the laser energy deposition $ E_{\mathrm{dep}} $ for $ \lambda_{\mathrm{laser}} = 1.064 $ $\mu$m (solid lines) and $ \lambda_{\mathrm{laser}} = 10.6 $ $\mu$m (dashed lines). Good agreement between the codes is observed for both laser wavelength cases. The blue, red and orange submissions employ a standard ray-tracing approach over the complete path of the laser. The submission shown in green transitions from a ray-tracing approach in the underdense plasma to a wave optics approach near the critical electron density. Here, the 1D Helmholtz equations are solved along the evanescent ray which propagates beyond the critical electron density, as seen in Fig.\,\ref{fig:Laser_energy_deposition}. The ``kink'' in the $ E_{\mathrm{dep}} $ profiles at $ x \approx 64 $ $\mu$m is a remnant of the temperature profile which transitions from the analytic form in Eq. (2) to a constant $ T_{e} = 3 $ eV profile for $ x > 64 $ $ \mu$m. For $ \lambda_{\mathrm{laser}} = 1.064 $ $ \mu$m, we note an approximate factor of two difference between the blue and green/orange submissions for $ x > 50 $ $\mu$m. This is attributed to differences in the Coulomb logarithms calculated by the codes (see next paragraph).

\begin{figure}
    \centering
\includegraphics[scale=1]{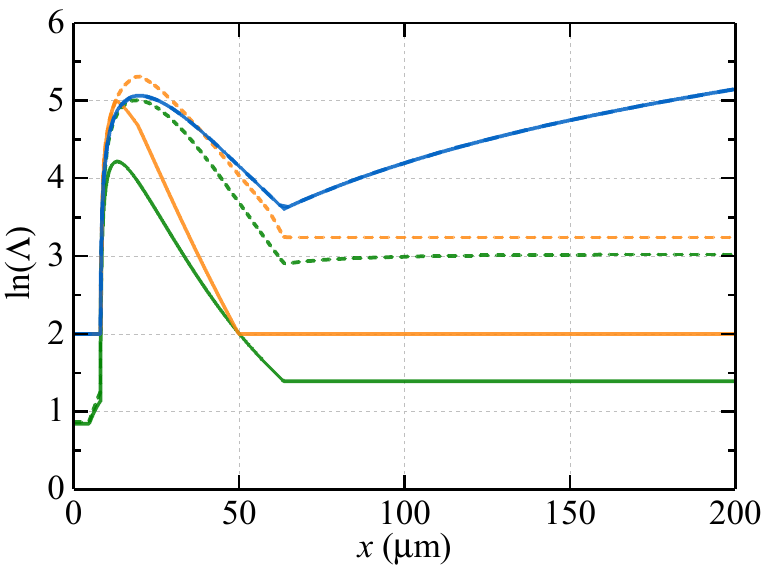}
    \caption{Coulomb logarithm $ \ln(\Lambda) $ as a function of distance $ x $. Solid lines correspond to the $ \lambda = 1.064 $ $ \mu $m case and the dashed lines correspond to the $ \lambda = 10.6 $ $ \mu$m case.}
    \label{fig:Coulomb_logarithm}
\end{figure}

The final two quantities we wish to discuss are the Coulomb logarithm $ \ln(\Lambda) $ and the laser absorption coefficient $ \alpha $. In Fig.\,\ref{fig:Coulomb_logarithm}, we plot $ \ln(\Lambda) $ as a function of distance $ x $ for the various codes. First, we note that the $ \lambda_{\mathrm{laser}} = 1.064 $ and 10.6 $ \mu$m cases overlap for the blue curve, implying that the approach used to derive $ \ln(\Lambda) $ is independent of laser wavelength. The $ \ln(\Lambda) $ value calculated by the blue submission is based on the formalism of Lee and More \cite{Lee1984}, where $ \ln(\Lambda) = \frac{1}{2}\ln(1 + b_{max}^2/b_{min}^2) $ and $ b_{max} $ ($ b_{min} $) are the maximum (minimum) impact parameters. In this approach, $ b_{max} = \mathrm{max}[\lambda_{DH},R_{0}] $, where $ \lambda_{DH} $ is the Debye-H{\"u}ckel screening length and $ R_{0} = (4\pi n_{i}/3)^{-1/3}$ is the average-ion radius. 

The calculation of $ \ln(\Lambda) $ for the submission shown in orange is based on the work of Skupsky \cite{Skupsky1987}. Unlike the model of Lee and More, the model of Skupsky explicitly accounts for the laser angular frequency $ \omega $ in the determination of $ b_{max} $ through $ b_{max} = \mathrm{min}[\mathrm{max}(\lambda_{DH},R_{0}),v_{t}/\omega] $, where $ v_{t} = (T_{e}/m_{e})^{1/2} $ and $ m_{e} $ is the electron mass. The value $ b_{max} = v_{t}/\omega $ corresponds to the high-frequency (low-density) plasma limit \cite{Skupsky1987}. While good agreement is observed between the blue and orange submissions for $ x < 64 $ $ \mu$m in the $ \lambda_{\mathrm{laser}} = 10.6 $ $\mu$m case (dashed curves), the orange submission exhibits a much steeper fall-off in $ \ln(\Lambda) $ in the $ 15 < x < 50 $ $\mu$m range for the $ \lambda_{\mathrm{laser}} =1.064 $ $\mu$m case (solid curves). Importantly, both models require $ \ln(\Lambda) \geq 2 $. 

Unlike the blue and orange submissions, the submission shown in green uses an interpolation formula for $ \ln(\Lambda) $, which enables a smooth transition from the weakly coupled plasma limit ($ \Lambda \gg 1$) to that of solid metals at room temperature ($ \Lambda \ll 1$). This approach yields good agreement with the orange submission, especially for the $ \lambda_{\mathrm{laser}} = 10.6 $ $\mu$m case. We note that the green submission predicts $ \ln(\Lambda) < 2 $ for $ x > 64 $ $\mu$m in the $ \lambda_{\mathrm{laser}} = 1.064 $ $\mu$m case.



\begin{figure}
    \centering
\includegraphics[scale=1]{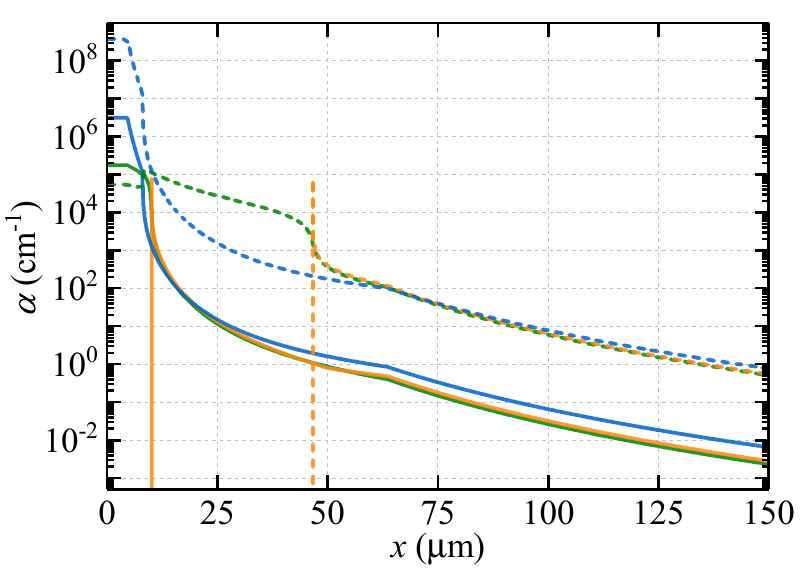}
    \caption{Laser absorption coefficient $ \alpha $ as a function of distance $ x $. Solid lines correspond to the $ \lambda = 1.064 $ $ \mu $m case and the dashed lines the $ \lambda = 10.6 $ $ \mu$m case.}
    \label{fig:Absorption_coefficient}
\end{figure}

Finally, we plot in Fig.\,\ref{fig:Absorption_coefficient} the laser absorption coefficient $ \alpha $ as a function of distance $ x $ for the $ \lambda_{\mathrm{laser}} = 1.064 $ $\mu$m and 10.6 $\mu$m laser cases. Good agreement between the codes in the underdense plasma regions is observed. The near factor-of-two difference between the green and blue submissions for the $ \lambda_{\mathrm{laser}} = 1.064 $ $\mu$m case is attributed to the aforementioned differences in the Coulomb logarithm. 

\section{Discussion and Outlook}

As motivated previously, the goal of this workshop was to initiate a code comparison activity in the EUV source plasma modelling community. In this respect, the workshop very much served its purpose. For one, the workshop highlighted a surprisingly large spread in predictions of the spectral purity of tin plasmas. This has its origins in the fact that (i) the codes predict different charge state distributions for a given $ T_{e} $ and (ii) the underlying atomic structures (from which the opacities and emissivities are built) can differ substantially from code-to-code. This has important consequences for radiation-hydrodynamic simulations of EUV sources, which take as input such radiative data to predict the conversion efficiency of laser-plasma EUV sources. These findings call for renewed investigations of tin-ion atomic structures and population kinetics in dense, laser-driven plasmas. Benchmark experiments of the ionization distributions of tin plasmas at EUV-generating conditions would greatly assist the validation and verification of collisional-radiative models.


As a follow-up to this first meeting, a 2$^{nd}$ EUV Light Sources Code Comparison Workshop was held on 25$^{th}$ October 2021 (the results of this meeting will be presented in a separate paper). A major development for this second meeting was that the software tools developed for the non-LTE workshops were made available to workshop organizers and participants. This enabled more detailed investigations and comparisons of a huge number of quantities (level populations, ionization rates, recombination rates, etc.) which were not investigated at the first workshop. The atomic kinetics problem was also expanded to study the effects of an external radiation field on the population kinetics. 

A new case study investigating radiation transport through a uniform tin sphere was defined. The goal of this problem was to obtain a self-consistent radiation field and material properties throughout the sphere. This problem served as an extension of the ``optically thin'' atomic kinetics case study, where optical depth effects play an important role in shaping the radiation field. Both steady-state and time-dependent variations of the problem were defined.

Finally, a time-dependent laser absorption case study was defined. In this problem, participants were asked to model the absorption of $ \lambda_{\mathrm{laser}} = 1.064 $ (Nd:YAG), 1.88 (Th) and 10.6 $\mu$m (CO$_{2}$) laser light in a one-dimensional planar tin plasma. As before, the plasma was assumed to be static (no hydrodynamic motion) and the processes of thermal conduction and radiation transport processes were to be omitted. In this way, the plasma could only gain energy through laser absorption and lose energy by radiating. The electron number density of the plasma was to be determined by evolving the ionization balance in time using non-LTE tin atomic kinetics. With this problem definition, we have edged closer to more ``realistic'' conditions whilst maintaining some degree of simplicity to ensure insightful comparisons.

\section{Conclusion}

In this paper, we have given an overview of the 1$^{st}$ EUV Light Sources Code Comparison Workshop. Two topics were addressed at the workshop. The first of these was an investigation of the atomic kinetics and radiative properties of tin plasmas at EUV-generating conditions. This case study highlighted a significant spread in predictions of the spectral purity of tin plasmas. This calls for renewed investigations of tin-ion atomic structures and plasma population kinetics processes. The second case study investigated laser absorption in a fully ionized, one-dimensional hydrogen plasma, where differences in the underlying Coulomb logarithms were found to be the principal source of disagreement among the codes.

\section{Acknowledgements}

We would like to thank all workshop participants, without whom this code comparison effort would not have been possible. We would also like to thank Dr. S. Langer and Dr. H. Frank for discussions during the preparatory stages of the workshop and Dr. Y. V. Ralchenko for making the non-LTE workshop software tools available for the 2$^{nd}$ EUV source code comparison workshop. Part of this work has been carried out at the Advanced Research Center for Nanolithography (ARCNL), a public-private partnership of the University of Amsterdam (UvA), the Vrije Universiteit Amsterdam (VU), NWO and the semi-conductor equipment manufacturer ASML.


\bibliography{mybibfile}

\end{document}